\documentclass
{amsart}
\usepackage{amssymb,amsmath,graphicx}
\theoremstyle{definition}

\theoremstyle{remark}

\numberwithin{equation}{section}

\begin{document}
\title [The Maxwell Equations] {An Asymmetry in the Maxwell Equations}
\author{J. Towe}
\address{Department of Physics, The Antelope Valley College, Lancaster, CA 93536}%
\email{jtowe@avc.edu}\
\begin{abstract}
A term in the Maxwell-Ampere law describes a linear displacement
current that is symmetrically enclosed by the curl of a magnetic
field. In this context symmetry calls for a term in the Faraday-Lenz
law, which in the absence of a conducting coil, would describe a
loop of displacement current about a linear increment of magnetic
flux. This term, introduced to satisfy symmetry, predicts a physical
phenomenon that has recently been observed.


\end{abstract} \maketitle
$ $\\[-06pt]
\section {The Faraday-Lenz and Maxwell-Ampere Laws}\label{S:intro}
The third Maxwell equation is
\begin{equation} \label{E:int}
\oint
\overrightarrow{E}\circ\overrightarrow{ds}=-\frac{d\Phi_{B}}{dt}.
\end{equation} This is the Faraday-Lenz law, which can be regarded
as describing an electric field that is induced along a conducting
coil by a changing magnetic flux that is symmetrically enclosed by
that coil. Due to the negative sign on the right side of Equation
1.1, the direction of the curl of the electric field is reversed.
The fourth Maxwell equation is
\begin{equation} \label{E:int}
\oint \overrightarrow{B}\circ
\overrightarrow{ds}=\mu_{0}\epsilon_{0}\frac{d\Phi_{E}}{dt}+\mu_{0}i_{enc},
\end{equation} where $\overrightarrow{B}$
represents a magnetic field and where the left side of the equation
is the integral of the dot product
$\overrightarrow{B}\circ\overrightarrow{ds}$ around a closed loop.
This is the Maxwell-Ampere law, which describes the right-handed
curl of a magnetic field about the linearly directed electric
current $\mu_{0}i_{enc}$, and about the displacement current
$\mu_{0}\epsilon_{0}\frac{d\Phi_{E}}{dt}$ (a virtual current that
occurs in the absence of a conductor and generates a magnetic
field). The constants of proportionality $\mu_{0}$ and
$\varepsilon_{0}$ are the familiar permeability and permittivity
constants. The term $\mu_{0}\epsilon_{0}\frac{d\Phi_{E}}{dt}$, which
describes Maxwell's virtual or displacement current was introduced
to satisfy symmetry, but was experimentally confirmed. 
$ $\\[-06pt]
\section {A New Term Dictated by Symmetry}\label{S:intro}
It is now argued that the above described equations collectively
involve an asymmetry that is due to the absence of a term on the
left side of the Faraday-Lenz law (1.1). This term would describe a
loop of displacement current about a linear increment
$\frac{d\Phi_{B}}{dt}$ of magnetic flux in the absence of a
conducting coil. According to conventional theory, the Faraday-Lenz
law would be degenerate in the absence of a conducting coil, but
symmetry and the existence of Maxwell's displacement current combine
to argue that a loop, or series of loops of virtual current (a loop
or series of loops of displacement current) does symmetrically
enclose a linear increment of magnetic flux in the absence of a
conducting coil. Specifically such considerations argue for the
addition of a term
\begin{equation} \label{E:int}
\oint \overrightarrow{E}_{virtual}\circ \overrightarrow{ds}
\end{equation} 
to the left side of the Faraday-Lenz law (Equation 1.1). 
As in the introduction of the term
\begin{equation} \label{E:int}
\mu_{0}\varepsilon_{0}\frac{d{\Phi_{E}}}{dt}
\end{equation} to the right side of the Maxwell-Ampere law, the
addition of the term (2.1) to the left side of the Faraday-Lenz law
is to satisfy symmetry; but as in the former case, the addition of
the term (2.1) predicts a phenomenon that can be experimentally
confirmed.
$ $\\[-06pt]
\section {Physical Prediction of the Proposed Term}\label{S:intro}
In paramagnetic materials such as water, each molecule is
characterized by a net permanent magnetic dipole moment. In the
presence of an external magnetic field, the magnetic dipole moments
of the individual molecules align with the external field. In this
context let us consider the molecules of a river that is flowing
from north to south. In this context the magnetic dipole moments of
the individual molecules align with the earth's magnetic field, so
that the paramagnetic sample of N molecules that is represented by
the river produces a magnetic dipole moment of magnitude $N\mu$
(where $\mu$ represents the magnetic dipole moment of an individual
molecule). Thus the north to south flow of the river produces an
effect that is equivalent to the motion of a large bar magnetic,
which is moving, north pole leading, toward the south. If the
proposed term (2.1), that (in the absence of a conducting coil)
describes a loop of displacement current about the time rate of
change of an actual magnetic flux, is correct, then the flow of the
river, which generates an increment $\frac{d\Phi_{B}}{dt}$ of
magnetic flux, should produce a loop of displacement current that is
orthogonal to the flow of the river; i.e. the flow of the river
should produce a curl of a virtual electric field,
$\overrightarrow{\nabla}X\overrightarrow{E}_{virtual}$, that is
anti-parallel to the flow of the river. In this context let us
consider a linear object consisting of a ferromagnetic material and
therefore associating with a weak magnetic field. If this
ferromagnetic object is prepared by exposure to the magnetic field
of the earth (so that the magnetic field that is generated by the
ferromagnetic material is aligned with the magnetic field of the
earth), and if this object is then exposed to the area above the
north-south flow of a river and oriented so that it lies in a plane
that is parallel to that of the flowing river, then this linear
object of ferromagnetic material should (if the fulcrum is
appropriately placed) experience a torque that tends to anti-align
the object with the north to south flow of the river. This
prediction was recently confirmed by D. Couillard.

\par
In the proposed context, the complete list of Maxwell's equations
is:
\begin{equation} \label{E:int}
\oint\overrightarrow{E}\circ
d\overrightarrow{A}=\frac{q_{enc}}{\epsilon_{0}}
\end{equation}
\begin{equation} \label{E:int}
\oint\overrightarrow{B}\circ d\overrightarrow{A}=0
\end{equation}
\begin{equation} \label{E:int}
\oint\overrightarrow{E\circ}d\overrightarrow{s}
+\oint\overrightarrow{E}_{virtual}\circ d\overrightarrow{s}
=-\frac{d\Phi_{B}}{dt}
\end{equation}
and
\begin{equation} \label{E:int}
\oint\overrightarrow{B}\circ\
d\overrightarrow{s}=\mu_{0}\epsilon_{0}\frac{d\Phi_{E}}{dt}+\mu_{0}i_{enc},
\end{equation}
where the proposed modification is described by the second term on
the left side of the Faraday-Lenz law (Equation 3.3). Again this
term describes a loop of displacement current in the absence of a
conducting coil, which is generated by a linear increment of
magnetic flux
$\frac{d\phi_{B}}{dt}$.  
\par

\end{document}